\documentclass{emulateapj}

\usepackage{natbib}
\usepackage[normalem]{ulem}
\shorttitle{4.8 keV Line in Mrk 876}
\shortauthors{Bottacini et al.}

\begin{document}

\title{An extreme gravitationally redshifted iron line at 4.8 keV in \object{Mrk~876}}
\author{Eugenio Bottacini\altaffilmark{1}, Elena Orlando\altaffilmark{1}, Jochen Greiner\altaffilmark{2},
Marco Ajello\altaffilmark{3}, Igor Moskalenko\altaffilmark{1}, \and Massimo Persic\altaffilmark{4,}\altaffilmark{5}}
\email{eugenio.bottacini@stanford.edu}

\altaffiltext{1}{W.W. Hansen Experimental Physics Laboratory \& Kavli Institute
for Particle Astrophysics and Cosmology, Stanford University, USA}
\altaffiltext{2}{Max-Planck-Institut f\"ur extraterrestrische Physik, Giessenbachstrasse 1, D-85748
Garching, Germany}
\altaffiltext{3}{Department of Physics and Astronomy, Clemson University, Clemson SC 29634-0978, USA}
\altaffiltext{4}{INAF-Trieste, via G.B.Tiepolo 11, I-34143 Trieste, Italy}
\altaffiltext{5}{INFN-Trieste, via A.Valerio 2, I-34127 Trieste, Italy}

\begin{abstract}
X--ray spectral lines at unforeseen energies are important because they can shed light on the extreme
physical conditions of the environment around the supermassive black holes of active galactic nuclei (AGN).
Mrk~876 displays such a line at $4.80^{+0.05}_{-0.04}$ rest--frame energy.
A possible interpretation of its origin can be found in the hotspot scenario. In this scenario
the primary radiation from a flare in the hot corona of an AGN illuminates a limited portion of the accretion disk that emits
by fluorescence. In this context the line can represent an extreme gravitationally redshifted Fe line
originating on the accretion disk below 6 gravitational radii from a rotating supermassive black hole.
The correct estimate of the line significance requires a dedicated approach. Based on an existing rigorous approach, we have 
performed extensive Monte Carlo simulations. We determine that the line is a real feature at $\sim$99\% confidence
level.
\end{abstract}


\section{Introduction}
Active galactic nuclei (AGN) are generous emitters of X-rays. These photons are thought to be
produced in the central region of the AGN. There, optical/UV photons from the accretion disk are
inverse-Comptonized by electrons (hot corona) thereby producing the primary X--ray radiation
seen by an observer as a power--law spectrum. This primary radiation also
irradiates the accretion disk that reflects it through several
reprocessing steps \citep[e.g.][]{matt91} including fluorescence emission of K$\alpha$ lines of most
abundant elements \citep{george91}. Therefore, emission lines from reflection spectra
are good diagnostics of the environment of the inner region of AGN
\citep[][and references therein]{ross93, fabian00}. Among emission lines 
from abundant elements in AGN spectra, the most common is the Fe K$\alpha$ line at 6.4 keV.\\ 
Even more intriguing are transient fluorescence emission lines. In this case the primary
radiation can originate in magnetic field reconnection \citep{merloni01} dissipating the energy input
into the Comptonizing corona and it irradiates a limited portion (hotspot) of the accretion disk for a short
time. As a result the emerging fluorescence lines are transient. They are also affected by Doppler and gravitational
effects causing distorted line shapes and photon energy shifts \citep{fabian00}. However, shifted line energies
can be caused also by AGN outflows and inflows. These shifts to unexpected
energies led to conflicting interpretations in the literature \citep{vaughan08} that are comprehensively discussed in \cite{turner10}.
Therefore, any new study is important.\\
In this Letter, we report on a detection in a {\em Swift}/XRT observation of an emission line
at 4.8 keV rest--frame energy of the optically discovered Seyfert type 1 AGN Mrk~876. We study the significance
of the line with Monte Carlo simulations. We then analyze all other available X--ray observations of the
source (by {\em GINGA}, {\em XMM-Newton}, and {\em Swift}/XRT) to constrain the physical origin of the line.

\section{Line feature in {\em Swift}/XRT spectrum of Mrk~876}
\begin{figure}[ht]
\epsscale{1.00}
\includegraphics[width=0.35\textwidth,angle=-90]{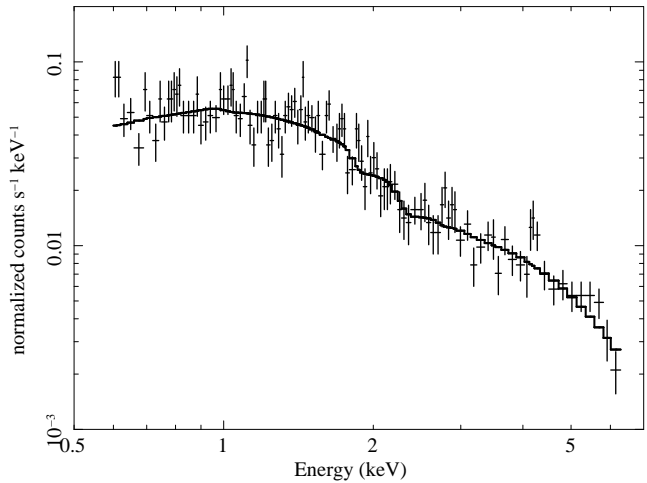}
\caption{{\em Swift}/XRT spectrum of Mrk~876 (observer's frame). Excess at 4.22 keV.}
\label{fig:spec}
\end{figure}

Mrk~876 is part of an optically selected sample of the Palomar-Green (PG) Bright Quasar Survey sample \citep{schmidt83}.
The redshift of the host galaxy is $z = 0.1385$ \citep{lavaux11} corresponding
to 551.4 Mpc for $H_{0} = 73$ km s$^{-1}$ Mpc$^{-1}$ assuming Hubble flow. The Galactic absorption toward
Mrk~876, assuming solar abundance, has been measured by \cite{elvis89} using NRAO 140 ft telescope of
Green Bank finding a value of N$_{H}^{gal}$ = 2.66 $\times$ 10$^{20}$ atoms cm$^{-2}$ with 5\% error.
The source was observed by {\em Swift}/XRT (obs id: 00050300004) on the 2005-03-01. 
For the analyses of the observation we used {\em HEAsoft 6.11.1} and {\texttt{xrtproducts}}. Events for the spectral
analysis are extracted from a circular region of interest centered on the source position having a
radius of $\sim$20 pixels \citep[corresponding to $\sim$47 arcsec,][]{moretti04}.
The background is extracted from a nearby source--free circular region of interest of 50 pixel radius.
The best fit model for the background-subtracted source spectrum is obtained by adding to the
power-law a Gaussian--component at 4.22 that models well the residuals (see Figure~\ref{fig:spec}).
All the parameters of the model are free to float obtaining a line energy of $4.22^{+0.05}_{-0.04}$ keV and
a width of $90^{+76}_{-75}$ eV compatible with the energy resolution of $\sim$~100 eV.
This leads to a fit result of $\chi^{2}=133.64/132~d.o.f$. The F--test shows strong evidence that the improved
fit result is not due by chance. 
To verify that the line is not due to the background, we analyze the background spectrum. It does not display
any line feature and it accounts for a mere 2.3\% of the total net count rate $1.409\times 10^{-1}$ count s$^{-1}$.
Furthermore its contribution to the line flux is negligible. The absence of spectral features and the low level of
background flux exclude any artifact that could give origin to the line. This is not surprising as Swift/XRT is known for
its low background level due to its low orbit especially for energies above 2 keV \citep{moretti07}.
In Section 2.1 we will estimate more properly the probability for the
Gaussian component to be a real spectral feature not caused by statistical fluctuations.
\subsection{Statistical significance of the line feature at 4.22 keV observed--frame energy}
\begin{figure}[ht]
\epsscale{1.00}
\centering
\includegraphics[width=0.50\textwidth,angle=0]{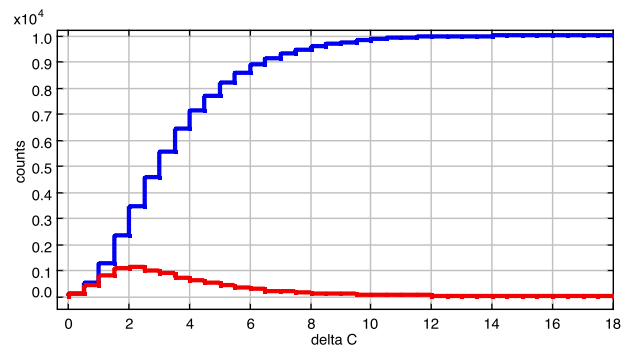}
\caption{Cumulative (blue) and frequency (red) histograms of delta C-statistic.}
\label{fig:cumulativechisquare}
\end{figure}

There is general consensus that line searches justified by observational
data must be in need of validation through randomized trials \citep{protassov02}.
\cite{turner10} used such an approach that we adopt with further improvements.\\
To estimate the significance of the line at 4.22 keV observed--frame energy, we carry out Monte Carlo simulations
in the wide energy range of 0.6 -- 6.5 keV. Our null hypothesis is: the spectrum as measured by {\em Swift}/XRT
is a power--law with absorption fixed to the galactic value (null model). This spectrum
is simulated by using the XSPEC command $fakeit$ accounting for the same exposure and instrument response files
of the observation. The simulated spectral data are grouped to a minimum of 15 counts per energy bin as done for
our measured spectrum thereby accounting for adequate statistics. This is iterated $10^{4}$ times. 
Each single simulated spectrum is fitted with our null model. The results are used to simulate a further
spectrum as before. This procedure accounts for the uncertainties of the null hypothesis model \citep{markowitz06, tombesi10}.
Again the obtained simulated spectrum is fitted with the null model thereby obtaining the $\chi^{2}_{null}$.
The same spectrum is then fitted with the null model and an additional new
Gaussian--line component throughout the 3.5 -- 6.5 keV energy range according to the work of
\cite{porquet04a} and successfully applied by \cite{markowitz06} and \cite{turner10}.
This energy range is stepped through by energy bins
whose size corresponds to the spectral energy resolution of {\em Swift}/XRT at each given energy \citep{short02}.
The line energy is allowed to freely float within this energy bin, similarly to what is done in a sophisticated line
search by \cite{turner10}.
Starting from an initial value of zero the line normalization freely varies between positive and negative values.
The line width is free starting from initial value of $\sigma$=0 to a maximum value corresponding to the energy
resolution. For each simulated spectrum the best chi square ($\chi^{2}_{best}$) is used to obtain the maximum
$\Delta\chi^{2}$ (= $\chi^{2}_{best}-\chi^{2}_{null}$). Consequently, the distribution of the $10^{4}$ $\Delta\chi^{2}$
indicates the fraction of lines caused by chance fluctuations whenever the value of the $\Delta\chi^{2}$ exceeds
the threshold value ($\Delta\chi^{2}_{thr}=8.77$).\\
To account for the low statistic domain we have performed further $10^{4}$ simulations as described above
but applying Churazov--weighting \citep{churazov96}. The $\Delta\chi^{2}$ need to exceed the
$\Delta\chi^{2}_{thr}=8.59$.\\
However, 15 counts bin$^{-1}$ might be low to confidentially use chi-square statistic as the probability
distribution might be Poisson \citep{cash79}. Therefore, to more confidentially estimate the line significance, we run
again $10^{4}$ simulations but applying $C$-statistic \citep{cash79} based on
maximum--likelihood ratio method. Even though $C$-statistic is suited for unbinned data, we group
spectra to 1 count bin$^{-1}$ avoiding zero--count bins as $C$-statistic is best performed by XSPEC for
minimum binning \citep{teng05}.
In this case $\Delta C$ (=$C_{best}-C_{null}$) must exceed $\Delta C_{thr}=10.95$.\\
As a result we find that thresholds of each of our 3 approaches chi-square, chi-square Churazov--weighted, and
$C$-statistic in each of their 10$^{4}$ spectra are exceeded by 1.36\%, 1.50\%, and 1.21\%, respectively.
These similar results are not surprising as differences between $\chi^{2}$ and $C$-statistic are expected when
number of counts bin$^{-1}$ and number of bins are low \citep{nousek89}, while here the latter is large.
Thus, the null hypothesis, of the measured spectrum being described by a power--law and galactic absorption,
is rejected by each of the 3 approaches with a probability of roughly 99\%. Figure~\ref{fig:cumulativechisquare}
displays the cumulative (blue) and the frequency histograms (red)
of the $\Delta C$ distribution: at $\sim$10.95 only a small fraction contributes to the total number of trials.\\
\subsection{Testing for systematic effects in the spectrum}
\begin{figure}[ht]
\centering
\includegraphics[width=0.45\textwidth, angle=0] {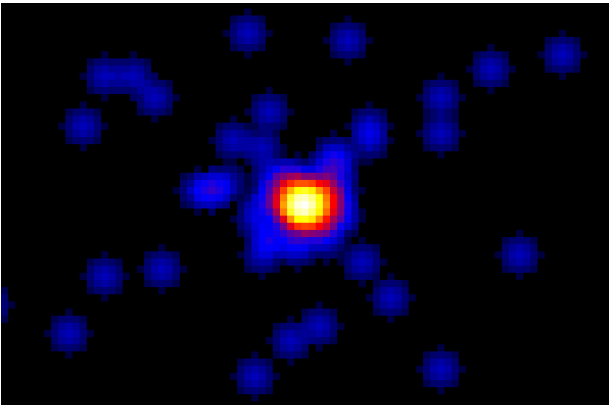}
\caption{{\em Swift}/XRT observation (00050300004) 4.0--4.4 keV.}
\label{fig:images-source}
\end{figure}

To further investigate on the reality of the line we undertake several tests. We extract spectra
from different positions on the sky (R.A., decl. = (243.3390, 65.6440), (243.2566, 65.5802),
(243.4108, 65.7710)) without detecting any spectral signature at 4.22 keV. This excludes systematics in this
specific observation. This is supported also by the fact that there has been no report on instrumental line features
at 4.22 keV due to calibration residuals of the effective area of {\em Swift}/XRT \citep{godet09}.
A further test is to extract the spectrum from Mrk~876 from nested and smaller circles of the source. The line feature
at 4.22 keV observed--frame energy still persists. We further verify that the line feature is not due to an
astronomical transient feature not associated to Mrk~876.
For this purpose, we first extract a light curve of Mrk~876 in the 4.0 -- 4.4 keV energy range from the observation
displaying the line (see Figure~\ref{fig:lc}). We find that the flux is stable over the entire observation time. Therefore,
a possible transient feature among our analyzed {\em Swift}/XRT observations not associated to Mrk~876 in the
image would be visible. Thus, we compare images in the 4.0 -- 4.4 keV energy band including the one with the
line feature (see Figure~\ref{fig:images-source}). There is no evidence for transient features.
These tests reinforce the idea that the line at 4.22 keV is not instrumental and that it is associated to Mrk~876.
\subsection{Further X--ray observations}
Mrk~876 was targeted by {\em Swift}/XRT with further 16 observations (see Table~\ref{tab:mrk876-fit}).
We neglect the 2 having less than 0.6 ksec exposure time and we analyze them in the same way as described earlier.
We find that the remaining spectra are best described by a power-law model with absorption fixed to the
Galactic value.\\
We have reanalyzed the {\em GINGA} \citep{makino87} spectrum of Mrk~876 as published in \cite{lawson97}. 
The 2 to 18 keV spectrum is best fitted with a power--law and absorption parameter fixed to the
Galactic value (see Table~\ref{tab:mrk876-fit}).\\
{\em XMM-Newton} has targeted the source twice (Table~\ref{tab:mrk876-fit}). These observations were analyzed
in \cite{piconcelli05} and in \cite{porquet04}. 
We have processed {\em XMM-Newton} Observation Data Files using the {\em XMM-Newton}
Scientific Analysis Software \citep{gabriel04} version 10.0.
Source spectra are extracted from a circular
region with radius of 30 arcsec while background spectra are extracted using a nearby source--free region
having 60 arcsec radius. The background--subtracted spectra are best fitted with a broken power--law and
absorption fixed to the Galactic value (see Table~\ref{tab:mrk876-fit}). There is no evidence for any line feature
including the Fe K$\alpha$ line at 6.4 keV.\\
Spectra were analyzed with XSPEC 12 \citep{arnaud96}.

\section{Discussions}
The importance of X--ray spectral lines at unforeseen energies is their capability to shed light on
the extreme physical conditions where they originate. Any new detection is important in light also
of future missions with improved throughput and energy resolution \citep[e.g. $Athena+$;][]{nandra13}.
\subsection{The line feature in Mrk~876}
Being a Seyfert type 1 AGN, Mrk~876 allows for clean view on the
innermost accretion region. As such, it is an ideal target for a case study.  None of the {\em GINGA}, {\em Swift}/XRT, and
{\em XMM-Newton} observations reveal the narrow Fe K$\alpha$ line at 6.4 keV. This is in agreement with previous
analyses of {\em GINGA} observation \citep{lawson97} and {\em XMM-Newton} observations \citep{porquet04, piconcelli05}.
This line, often seen in AGN, is interpreted as Fe fluorescence from cold (neutral) matter far from the inner accretion disk
\citep{yaqoob04, page04}. A viable explanation for its absence in Mrk~876 can come from the X--ray Baldwin effect
\citep{iwasawa93}, an anti-correlation between the equivalent width of the Fe K line and X--ray luminosity, which can
be explained by the drop of the covering factor of the molecular AGN torus \citep{page04}.
Therefore, it is not surprising that Mrk~876 does not display a steady Fe K$\alpha$ line being one of the most luminous
AGN among the systematically  studied Palomar-Green (PG) Bright Quasar Survey sample at X--rays \citep{porquet04}. 
However, for one {\em Swift}/XRT observation (obs id: 00050300004) we find a significant line emission at a confidence
level of 99\% at $4.80^{+0.05}_{-0.04}$ keV rest--frame energy. At the same rest--frame energy a similar line but with
lower confidence level has been reported by \cite{petrucci07} for Mrk~841 that is a Seyfert type 1 AGN.
\subsection{On the origin of the 4.8 keV rest--frame line}
\begin{figure}[ht]
\centering
\includegraphics[width=0.50\textwidth, angle=0] {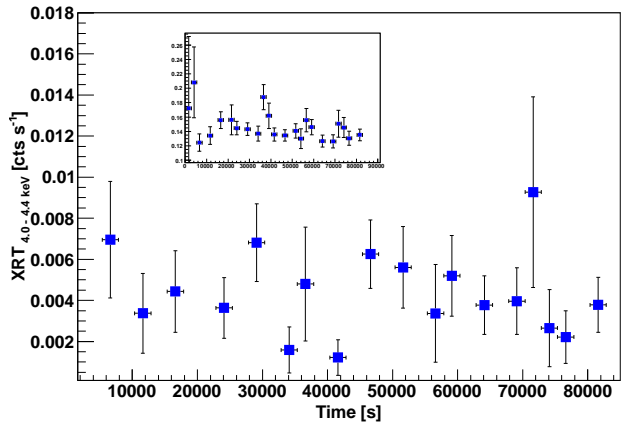}
\caption{Light curve of observation 00050300004 in the energy band 4.0 -- 4.4 keV and 0.6 -- 6.5 keV (inset).}
\label{fig:lc}
\end{figure}

In the following we consider the possible origin of our transient line in the accretion disk hotspot scenario
\citep{nayakshin01, turner02}. The inclination angle of Mrk~876's accretion disk of $15.4^{+12.1}_{-6.8}$
deg is independently constrained by \cite{bian02}.
This very low inclination angle leads to the gravitational redshift dominance of the line over the Doppler effect in both rotating 
(Kerr) black hole \citep{laor91} and a non-rotating (Schwarzschild) black hole environments \citep{fabian89}. As a result the
line profile tends to a single peak \citep{fabian89}, which is intrinsically very narrow \citep[$<$100 eV;][]{nayakshin01}
and mainly subject to gravitational redshift.
The observed gravitational redshift factor is $\nu_{observed}/\nu_{restframe}$ = $0.75^{+0.01}_{-0.01}$ where errors account
for the line energy uncertainty. Following \cite{fabian89} this factor cannot be obtained by a non-rotating black hole.
We constrain the emission-region on the accretion disk by simulating the model of \cite{laor91} accounting for its different emissivity
power-law indices and taking care of the uncertainties of the inclination angle of the source. The gravitational
redshift factor of $0.75^{+0.01}_{-0.01}$ cannot be obtained from emission-regions at radii above $6~r_{g}$
(where $r_{g}=GM_{BH}/c^{2}$) reaching at most a factor of 0.89. On the contrary for radii below $6~r_{g}$
such a gravitational redshift factor can be well reproduced. Only accretion disks of Kerr black
holes can extend below $6~r_{g}$ \citep{thorne74} emphasizing once again the environment of a rotating black hole. On the other hand, for Schwarzschild black holes, a detailed
line profile study by \cite{reynolds97} allows for the fluorescence Fe K$\alpha$ line emission from material spiraling
toward the event horizon even below 6 $r_{g}$. However, the profile is predicted to be double--peaked for accretion
disk inclination angles other than face--on, in contrast to our finding.
As a result, the line can be reproduced by a rotating black hole only.
This is in agreement with a
range of spectral parameters predicted in an accurate study by \cite{dovciak04} for a black hole with angular momentum
$a$=0.9 and disk inclination $\theta_{0}$=30 deg for a hotspot on an accretion disk emitting photons from its last stable orbit
$r_{ms}+dr$.
Unfortunately the low signal-to-noise spectrum does not allow for a more constraining fit with the Laor--model or with the more complex {\texttt{kerrdisk}}-model.
The same low
signal-to-noise ratio prevents from attempting a cross-correlation study between the line light curve and the continuum
light curve (both shown in Figure~\ref{fig:lc}) that are predicted to correlate in the hotspot framework \citep{dovciak04}. 
Also correlation studies between {\em Swift}/XRT and {\em Swift}/BAT are impossible due to the low signal-to-noise
observations of the latter \citep{baumgartner13}. Unfortunately, no UVOT data were taken, which could constrain the
geometry of the emitting region \citep{george91}. On the other hand the independent estimates of the black hole mass
of Mrk~876 \citep{kaspi00, bian02} allow us to compute the orbital period for the hotspot on the accretion disk.
\cite{dovciak04} predict the line to appear for a relatively short fraction of the total orbital period of the hotspot at the
distance from the black hole where the line occurs. Thus, we estimate the orbital period using the formula below
of \cite{bardeen72} as performed in \cite{dovciak04} where $r$ is in units of gravitational radii.
\begin{equation}
P = 310~\left(r^\frac{3}{2}+a\right)
\frac{M_{BH}}{10^7M_{\odot}}\quad\mbox{[s]}.
\label{torb}
\end{equation}
We assume the photons emitted from a spinning black hole $a$=0.9 at a disk radius of
$6~r_{g}$. Taking the two independently inferred black hole
masses of Mrk~876 $M_{BH}$=1.3$\times$10$^{9}$ M$_{\odot}$ \citep{bian02} and
$M_{BH}$=2.4$\times$10$^{8}$ M$_{\odot}$ \citep{kaspi00}, the resulting
orbital periods are P=628 ksec and P=135 ksec, respectively. Both are significantly longer than the {\em Swift}/XRT
observation time ($\sim$25 ksec). This is in agreement with the narrow lines to appear for a fraction of the total period
as the flare dies out or emission gets below detectability \citep{dovciak04}.
The {\em Swift/XRT} observation (obs id: 00050302001) taken 6 days later without
any line detection, makes the hotspot scenario a consistent picture. Also the line feature has never been observed in
any other X--ray observation confirming the transient nature of the event that reinforces the hotspot origin.
\\
\\
Another hypothetical origin of the line could be the emission from enhanced abundances of sub--Fe elements
resulting from Fe spallation.
In this scenario, Fe nuclei hit by cosmic--rays are converted into elements of atomic number $Z$
below Fe. The element emitting closest to our line energy would be Vanadium at 4.9 keV (rest--frame energy).
However, the absence of the Fe K$\alpha$ at 6.4 keV, which would still be the strongest line in this scenario
\citep{skibo97}, makes it an unlikely explanation. Also the transient nature of the line would be difficult to explain.\\
In a further hypothesis a $\sim$4.7 keV (rest--frame energy) spectral feature could originate also from Compton
down-scattered Fe line from FeXXV at 6.7 keV \citep{fabian12}. However, the resulting gaussian component is
expected to be much broader ($\sim$ 0.4 keV). Also, this component needed
to be present in all our analyzed Mrk~876 spectra. Thus, the down-scattered Fe line is also very unlikely.\\
A redshifted Fe line could originate also from inflows and outflows of material (not related to the accretion
disk) in the vicinity of the black hole. In this region the material would be subject to strong gravitational effects 
reaching mildly relativistic ($\sim$0.4c) velocities \citep{tombesi10} causing an energy shift of emission lines.
This is the interpretation for a line shift observed in Mrk~766 in a detailed and time-resolved study by \cite{turner04}
in presence of a strong Fe K$\alpha$ component at 6.4 keV. In fact, the absence of this latter component in all
our analyzed observations of Mrk~876 disfavors this interpretation since it is difficult to imagine one single blob
of material emitting an Fe line. In addition our observations a month earlier (obs id: 00050300005) and 6 days
later (obs id: 00050302001) with {\em Swift}/XRT should have detected the corresponding 6.4 keV line.

\subsection{On the significance of the line}
A detailed study by \cite{vaughan08} suggests that reported line significances might be
altered by a publication bias where only most significant among detected spectral features are
mentioned. Such an important issue can be addressed by a statistical study of a sample as a whole
where several lower significance lines are detected as e.g. in \cite{tombesi10}. Instead, our study reports a
case study of a transient event in the hotspot scenario. Therefore, to determine the detection likelihood we
cannot incorporate the non-presence of the lines in other observations. Also is it here impossible to address
the publication bias as it must address the bias for a source population at large \citep{vaughan08}.

\section{Conclusions}
Our analysis of the {\em Swift}/XRT observation of Mrk~876 (obs id: 00050300004) has revealed an emission
line at $4.80^{+0.05}_{-0.04}$ keV rest--frame energy associated to the source. The line is detected at $\sim$99\%
confidence level through Monte Carlo simulations using $\chi^{2}$ and $C$-statistic. The possible origin of the
line is the hotspot scenario in which a flare in the hot corona (possibly due to magnetic reconnection) illuminates
a limited portion of the accretion disk that emits the Fe line by fluorescence. Due to the independent constraints of
the low inclination of Mrk~876's accretion disk (15.4 deg) and the observed gravitational redshift, we can constrain
that the line is emitted from the accretion disk at radii below 6 $r_{g}$ (where $r_{g}~=~GM_{BH}/c^{2}$) from a
Kerr black hole. The alternative possible origins due to Fe spallation, down-scattered Fe line, and mass
inflows/outflows are disfavored in light of the absence of the 6.4 keV Fe line and the variability studies due to the
many available observations of {\em Swift}/XRT, {\em XMM-Newton}, and {\em GINGA}.\\
To the best of our knowledge, this is the first time that a transient line revealed at 99\% confidence level is
unambiguously associated to the hotspot scenario.

\acknowledgments
Considerable improvement of the Letter is due to robust comments of the anonymous referee.
E.B. acknowledges NASA grant NNX13AF13G. The authors thank Giacomo Vianello for helpful
discussions and the instrument teams for the observations.

{\it Facilities:} \facility{Swift}, \facility{XMM}, \facility{Ginga}

\clearpage

\begin{turnpage}
\begin{deluxetable*}{cccccccccc}
\tablewidth{0pt}
\tabletypesize{\scriptsize}
\tablecaption{Mrk 876 soft X-ray spectral fit. $^{*}$observation displaying the line. \label{tab:mrk876-fit}}
\tablehead
{
\colhead{instrument} 			&\colhead{start}				&
\colhead{expo} 				&
\colhead{$\Gamma$}    			& \colhead{$\Gamma_{hard}$}     	&
\colhead{E$_{break}$} 			& \colhead{norm} 				&
\colhead{Chi-square} 			&\colhead{d.o.f.} 				&
 \colhead{flux 2-6 keV} \\
\colhead{\scriptsize [$obs~id$]} 			& \colhead{\scriptsize[$date$]}			&
\colhead{\scriptsize[$sec$]}				&
\colhead{\scriptsize}		 	& \colhead{} 			&
\colhead{\scriptsize[$keV$]}                       	& \colhead{[$10^{-3}~ph~keV^{-1}~cm^{-2}$]} 	& 
\colhead{} 						&\colhead{}				&
 \colhead{\scriptsize[$10^{-12}~erg~cm^{-2}~s^{-1}$]}
}
\startdata
GINGA  & 1991-05-11 12:10:00 & 112208 & 1.47$^{+0.10}_{-0.11}$ &  \nodata &  \nodata  & 0.8$^{+0.2}_{-0.1}$ & 12.44 & 11 & 2.97$^{+0.25}_{-0.27}$ \\
XMM 0102040601 & 2001-04-13 19:38:25 & 12825 & 2.50$^{+0.13}_{-0.15}$ &  1.81$^{+0.12}_{-0.22}$ &  0.97$^{+0.45}_{-0.19}$ & 1.01$^{-0.21}_{-0.10}$ & 149.77 & 136 & 2.43$^{+0.20}_{-0.17}$ \\
XMM 0102041301 & 2001-08-29 07:45:05 & 7919 & 2.48$^{+0.06}_{-0.07}$ &  1.76$^{+0.15}_{-0.14}$ &  1.46$^{+0.24}_{-0.34}$ & 1.85$^{+0.07}_{-0.10}$ & 184.10 & 200 & 3.30$^{+0.18}_{-0.13}$ \\
XRT 00050300001 & 2004-12-23 00:12:23 & 3774 & 1.12$^{+0.17}_{-0.16}$ &  \nodata &  \nodata & 0.59$^{+0.09}_{-0.09}$ & 14.08 & 12 & 3.20$^{+0.32}_{-0.34}$ \\
XRT 00050300005 & 2005-01-26 00:37:41 & 12669 & 1.26$^{+0.06}_{-0.07}$ &  \nodata &  \nodata & 0.94$^{+0.06}_{-0.05}$ & 117.45 & 105 & 4.28$^{+0.18}_{-0.20}$ \\
XRT 00050300004$^{*}$ & 2005-03-01 23:59:01 & 25458 & 1.50$^{+0.06}_{-0.05}$ &  \nodata &  \nodata & 0.77$^{+0.03}_{-0.04}$ & 142.41 & 135 & 2.52$^{+0.09}_{-0.10}$ \\
XRT 00050302001 & 2005-03-08 02:08:54 & 23158 & 1.36$^{+0.05}_{-0.05}$ &  \nodata &  \nodata & 0.87$^{+0.04}_{-0.04}$ & 169.76 & 153 & 3.42$^{+0.11}_{-0.12}$ \\
XRT 00050302002 & 2005-03-09 00:33:52 & 9400 & 1.37$^{+0.09}_{-0.08}$ &  \nodata &  \nodata & 0.87$^{+0.06}_{-0.07}$ & 72.19 & 69 & 3.37$^{+0.17}_{-0.18}$ \\
XRT 00050303001 & 2005-03-09 08:55:52 & 7725 & 1.27$^{+0.10}_{-0.09}$ &  \nodata &  \nodata & 0.83$^{+0.07}_{-0.07}$ & 49.32 & 53 & 3.67$^{+0.23}_{-0.23}$ \\
XRT 00050303003 & 2005-03-12 00:50:03 & 3879 & 1.35$^{+0.12}_{-0.12}$ &  \nodata &  \nodata & 0.92$^{+0.10}_{-0.11}$ & 23.71 & 31 & 3.67$^{+0.29}_{-0.32}$ \\
XRT 00050303004 & 2005-03-24 07:07:51 & 10269 & 1.42$^{+0.08}_{-0.08}$ &  \nodata &  \nodata & 0.86$^{+0.05}_{-0.07}$ & 94.27 & 71 & 3.12$^{+0.18}_{-0.17}$ \\
XRT 00035308001 & 2006-05-30 07:31:01 & 8463 & 1.76$^{+0.08}_{-0.09}$ &  \nodata &  \nodata & 1.41$^{+0.08}_{-0.09}$ & 76.01 & 77 & 3.32$^{+0.15}_{-0.18}$ \\
XRT 00035308002 & 2006-06-18 01:12:01 & 4422 & 1.77$^{+0.14}_{-0.14}$ &  \nodata &  \nodata & 1.25$^{+0.12}_{-0.14}$ &40.36 & 30 & 2.90$^{+0.23}_{-0.26}$ \\
XRT 00035308003 & 2007-09-25 00:11:00 & 2214 & 1.83$^{+0.19}_{-0.18}$ &  \nodata &  \nodata & 1.28$^{+0.17}_{-0.15}$ & 4.94 & 12 & 2.78$^{+0.32}_{-0.31}$ \\
XRT 00035308004 & 2007-10-02 17:02:00 & 2590 & 1.69$^{+0.15}_{-0.14}$ &  \nodata &  \nodata & 1.16$^{+0.12}_{-0.14}$ & 17.16 & 21 & 2.97$^{+0.28}_{-0.30}$ \\
XRT 00035308005 & 2007-10-04 04:16:00 & 2887 & 1.96$^{+0.16}_{-0.16}$ &  \nodata &  \nodata & 1.63$^{+0.17}_{-0.16}$ & 11.58 & 19 & 3.01$^{+0.30}_{-0.27}$ \\
XRT 00035308006 & 2007-10-04 10:35:00 & 5927 & 1.81$^{+0.11}_{-0.10}$ &  \nodata &  \nodata & 1.37$^{+0.11}_{-0.10}$ & 41.06 & 51 & 3.03$^{+0.19}_{-0.21}$ \\
XRT 00091642002 & 2013-10-02 13:21:59 & 1012 & 1.66$^{+0.35}_{-0.34}$ &  \nodata &  \nodata & 2.13$^{+0.45}_{-0.46}$ & 7.97 & 10 & 5.64$^{+0.74}_{-0.94}$ \\
\enddata
\end{deluxetable*}
\end{turnpage}

\end{document}